\documentclass[12pt]{article}
\usepackage[dvips]{color}
\usepackage[dvips]{graphics}
\usepackage{a4}

\begin{document}

\title{DETERMINISM AND THE GRAVITATIONAL PLANCK CONSTANT.}

\author{Antonio Zecca \\
Istituto Nazionale per la Fisica della Materia, \\
Dipartimento di Fisica, Universit\`a di Trento. \\
Trento - Italy
}

\maketitle

\begin{abstract}

A discussion is given of the uncertainty principle in view of the
introduction of a Gravitational Planck Constant.  The need for such a
gravitational constant is shown first.  A reduced electromagnetic Planck
constant and the analogous reduced gravitational Planck constant are defined
as $h/e^2$ and $H/m^2$ respectively.
An attempt is made to reconcile the quantum uncertainty concepts with a
deterministic view of the physical world.
This conclusion is achieved trough the detailed analysis of the measurement
procedures of physical quantities.
 
\end{abstract}
 
 
\newpage

\section{INTRODUCTION.}
 
A single value of the Planck constant has been used until now when treating
both electromagnetic and gravitational phenomena.  This is a reasonable
assumption since most of the gravitational experiments we can think about are
performed using simultaneously electromagnetism and gravitation: all our
detectors are electromagnetic devices.  Neverthless it is easy to think about
phenomena where electromagnetism does not play a role.  It has been proposed
that when dealing with \textit{purely gravitational }phenomena, the Planck
constant $h$ should be replaced with a gravitational counterpart ($H$ from now
on)~\cite{ref1,ref2}.  $H$ has been defined as~\cite{ref1}:
\begin{equation}
H = \frac{1}{8 \pi~\varepsilon_0~G} \times
\frac{m^2}{e^2} \times h
\label{eq1}
\end{equation}
with a value: $H = 7.95 \times 10^{-77}$~J s. Here $e$ and $m$ are the
electron charge and mass and therefore this constant relates to the
electron. This aspect will be discussed later.
 
\section{THE NEED FOR A GRAVITATIONAL PLANCK CONSTANT.}
 
In order to evidence the need for a Gravitational Planck Constant, we will
start with a discussion of the emission mechanism from an electromagnetic
dipole, a very well understood case.  In such a system the quantum theory
holds that photons are emitted with an energy $E = h \nu$.  A 1000~MHz
microwave antenna with a RF power of 10~W is supposed to emit some $10^{25}$
photons per second.  Each photon is admitted to show a particle-like
behaviour, while the ensemble behaviour of these photons gives rise to the
wave aspect of their propagation.  The particle-like behaviour of each photon
is not accessible to us in this case, only because of the technological limits
of our detectors~\cite{ref3}.  We observe waves, but definitely nobody would
conclude that the emission is not quantized as photons.  The number of photons
emitted per second can be easily evaluated starting from realistic numbers:
for instance a 10~W electronics creating a 1~A RF current from a 10~V supply.
In this case some $6 \times 10^{18}$ electrons are accelerated through a 10~V
potential difference every second.  We can divide the 10~eV associated energy
drop into steps of size $h \nu / e$ ($\simeq 4 \times 10^{-6}$~V at 1000~Mhz),
getting $2.5 \times 10^6$ such "energy levels".  It is within the quantum
ideas to think of a photon being created every time one electron passes such
steps.  We end up with $\simeq 10^{25}$ photons per second.  This calculation
is obviously tautological, but has the purpose of evidencing the physical
process.

We do not expect gravitational (dipole) radiation to arise from such a
process.

Neverthless, starting from the above example, we can easily accept that in
more complex systems (like an exploding supernova or a pulsar) the emitting
agents are the single elementary particles involved in the macroscopic process
- anytime a quadrupole change in the gravitational field acts on them.  This
view has been proposed earlier: see for instance Ref.~\cite{ref4}.  The
correct interpretation of quantum ideas suggests that a large number of
gravitational quanta are emitted: in analogy with the RF dipole example we
expect to observe a gravitational wave.  Being the gravitational field more
than 40 orders of magnitude weaker than the electromagnetic field involved
during the same processes, we expect that the gravitational wave carries an
energy some 40 orders of magnitude smaller than the electromagnetic one. If we
want to keep the energy conservation law, we have two choices:
 
\begin{enumerate}
\item{hipothesize that the emission probability of each elementary emitter is
more than 40 orders of magnitude smaller for gravitational than for
electromagnetic emission (see for instance Zeldovich and Novikov~\cite{ref4});
or}
\item{hipothesize that the energy carried by each graviton is more than 40
orders of magnitude smaller than the energy of the corresponding photon, the
emission probability being of the same order of magnitude.}
\end{enumerate}
 
The first possibility is rather queer, and it does not comply with the
above description of the emission process; in addition, it has never been
proved. The second possibility calls for the use of a gravitational
Planck constant as defined in Eq.~(\ref{eq1}).
 
According to such definition, this constant depends on the mass of the test
particles involved in the gravitational process.  This peculiarity is hardly
surprising if we consider that charge takes only one value as it does the
electromagnetic Planck constant $h$, while particles appear with a spectrum of
masses. Therefore, the use of a mass-dependent gravitational constant is the
natural extension of the single valued electromagnetic one.
 
\section{THE REDUCED PLANCK CONSTANTS.}

Neverthless it has a meaning the definition of a "reduced gravitational Planck
constant":
\begin{equation}
H^* = H / m^2
\label{eq2}
\end{equation}
which is independent on the mass of the emitting system and the anologous
"reduced electromagnetic Planck constant": 
\begin{equation}
h^* = h / e^2
\label{eq3}
\end{equation}
with numerical values:
\begin{eqnarray*}
H^* & = & 9.58 \times 10^{-15}~\rm{J~s / kg}^2 \\
h^* & = & 2.58 \times 10^{4}~\rm{J~s / coul}^2
\end{eqnarray*}

We note here that the reduced electromagnetic Planck constant enters the
equations governing several physical phenomena (the Bohr formula, the fine
structure constant, the relativistic correction to the hydrogen atom energy
levels, the quantum Hall effect - for example).  Alfonso-Faus has proposed a
new system of units such that Local Lorentz Invariance and Local Position
Invariance is preserved~\cite{ref5}. In this system
$\varepsilon_0 = \mu_0 = 1/c$
and the fine structure constant turns out to be the inverse of the reduced
electromagnetic Planck constant:
$$
\alpha = \frac{e^2}{2 h} = \frac{1}{2 h^*}
$$

The reduced electromagnetic Planck constant can be explicited also in the
relation
\begin{equation}
E = h \nu = h^* \times e^2 \times \nu
\label{eq:4}
\end{equation}

This writing has the advantage of showing explicitely the dependence of the
energy of the electromagnetic quantum on the value of the emitting charge,
besides the well known  dependence on the frequency.  The charge dependence is
usually ignored because at the microscopic level charge takes a single value
in most cases. 

Coming back to the multiple valuedness of the gravitational Planck constant
(Eq.~\ref{eq1}), we can compare Eq.~(\ref{eq:4}) with the gravitational
analogue:
\begin{equation}
H = H \nu = H^* \times m^2 \times \nu
\label{eq:5}
\end{equation}

We see that this  multiple valuedness implies that different particles
undergoing the same energetic transition will emit gravitational quanta with
different frequencies.

\section{DETERMINISM OR UNCERTAINTY?}
 
The introduction of a gravitational Planck constant prompts the idea of
"gravitational uncertainty relations"~\cite{ref1,ref2}. For instance, in
\textit{purely gravitational} phenomena the momentum uncertainty relation
should be written:
\begin{equation}
\Delta x ~ \times ~ \Delta p_x = \frac{H}{2 \pi}
\label{eq:6}
\end{equation}
 
We would be faced now with two levels of uncertainty: the "coarse"
electromagnetic level and a much finer gravitational level.  Are the two in
contraddiction? Do they pose unsolvable inconsistencies?  Can we violate the
limits of the electromagnetic uncertainty by taking advantage of the
gravitational one?  In this paragraph we will give a tentative answer.
 
Any observation of microscopic phenomena, must use a suitable probe.  A
careful analysis of the fundations of our experimental techniques shows that
our probes all satisfy two requirements.  First, the energy of the probe is
always of (it cannot be other than) the same order of magnitude of the typical
energies involved in the system under observation.  Second, all probes are -
directly or indirectly - electromagnetic in their nature.  These statements
are obvious in a sense, since anybody knows implicitely that they are true;
neverthless, at our knowledge, they have not been stressed in the analysis of
the uncertainty concepts.  The consequence of the mentioned requirements is
that as soon as we apply our probe to the measured system, we perturb it - as
it is well known. This is different from our macroscopic experience, where we
can choose the probe such as to reduce the perturbation (the "measurement
error") to an acceptable (say sub-one-percent) level.  At the microscopic
level, if we prepare a particle to travel in a certain direction, it will
continue to travel in the same direction with the same momentum, unless we
shoot it with a probe.  This is guaranteed by the inertia principle, which is
believed to be valid at all levels in physics.  In an experimental set-up
aimed at the test of Eq.~(\ref{eq:6}), in order to give a measurement result
we shall prepare a large number of target particles - a beam - and we shall
shoot them with a large number of probe particles: we need such large numbers
to achieve a statistically significant figure as the result of our
experiment. After the interaction, we will observe a distribution of momenta
for the outgoing target particles.  This distribution reflects the
distribution of interaction channels which have been sampled by different
target-probe couples. We conclude that the probabilistic behaviour is not
intrinsic to the observed system, but is caused by our measurement protocol
and by the rudeness of our probes.  The simplest and plainest description of
such a measurement process requires that the behaviour of the target particles
is deterministic, while the results of our observation are (cannot be other
than) probabilistic.

\section{CONCLUSIONS}
 
Coming back to the gravitational uncertainty relations, it is obvious that
\textit{purely} gravitational probes offer in principle a reduced perturbation
on the measured system. The smaller intensity of the force gives a support to
the use of the gravitational Planck constant.  A much smaller uncertainty
could be possible in our measurements, provided we could get the technology
for the production and detection of single gravitons.  Lacking such devices
and using instead gravitational wave detectors we would be very much in the
situation of the 19th century wave optics: no need for quantization.  On a
different point of view, single graviton effects would be buried by far in the
electromagnetic thermal noise in all practical situations. From the conceptual
point of view the introduction of the gravitational relations will not create
a new limit of uncertainty; based on the above analysis, the physical world
remains deterministic: only the results of our measurements would be
represented with different statistical distributions.
 
\newpage

\noindent\textbf{ACKNOWLEDGEMENTS}

\smallskip

The author thanks Prof. Luciano Vanzo for useful discussions.


\begin{thebibliography}{1}
\bibitem{ref3}
At our knowledge no single photon detection in this frequency range ($10^3$ to
$10^5$~MHz) has been performed until now. Neverthless the technology for such
experiments is available since cooling at mK temperatures is possible with
commercial equipment.
\end{thebibliography}
\end{document}